\documentclass{article}[12pt]
\usepackage{amssymb, amsthm, amsmath}
\usepackage[a4paper]{geometry}
\usepackage{amssymb,latexsym,amscd}
\usepackage{epsfig}
\usepackage[active]{srcltx}
\usepackage{psfrag}
\usepackage{graphics,graphicx}
\usepackage{pstricks,pst-node,pst-tree}
\usepackage{braket}
\usepackage{color}
\usepackage{array}
\usepackage{booktabs}

\usepackage{natbib}

\usepackage{tikz}
\usetikzlibrary{arrows.meta,calc,decorations.markings,math,arrows.meta}
\usetikzlibrary{backgrounds}
\DeclareMathOperator\likelihood{likelihood}

\usepackage{bm}
\usepackage{hyperref}

\usepackage{caption}
\usepackage{subcaption}

\title{Distinguishing between convergent evolution and violation of the molecular clock}
\author{Jonathan D. Mitchell$^{1,2}$, Jeremy G. Sumner$^1$, and Barbara R. Holland$^1$}

\begin{document}

\maketitle

\begin{center}
\noindent {\small \it 
$^1$School of Physical Sciences, University of Tasmania, Hobart, Tasmania 7001, Australia} \\
\noindent {\small \it 
$^2$Department of Mathematics \& Statistics, University of Alaska Fairbanks, Alaska 99775, USA}
\end{center}
\medskip
\noindent E-mail: jdmitchell5@alaska.edu.\\

\begin{abstract} 
We give a non-technical introduction to convergence-divergence models, a new modeling approach for phylogenetic data that allows for the usual divergence of species post speciation but also allows for species to converge, i.e. become more similar over time. 
By examining the $3$-taxon case in some detail we illustrate that phylogeneticists have been  ``spoiled''  in the sense of not having to think about the structural parameters in their models by virtue of the strong assumption that evolution is treelike. 
We show that there are not always good statistical reasons to prefer the usual class of treelike models over more general convergence-divergence models. 
Specifically we show many $3$-taxon datasets can be equally well explained by supposing violation of the molecular clock due to change in the rate of evolution along different edges, or by keeping the assumption of a constant rate of evolution but instead assuming that evolution is not a purely divergent process. 
Given the abundance of evidence that evolution is not strictly treelike, our discussion is an illustration that as phylogeneticists we often need to think clearly about the structural form of the models we use.
\end{abstract}

\noindent (Keywords: convergence-divergence models, molecular clock, phylogeny, identifiability, Akaike information criterion)\\

\section{Modeling Evolution With Convergence-Divergence Models}

In \citet{sumner2012algebra}, the authors introduced a new model class that generalizes the standard Markov model of character evolution on a phylogenetic tree. 
The formulation is similar to the idea of an $n$-taxon process \citep{bryant2009hadamard}, where the Markov process acts on the state space of all possible character patterns for $n$ taxa.
However, Bryant's $n$-taxon process is restricted to the Kimura 3ST model of sequence evolution, whereas the results given in \citet{sumner2012algebra} are valid for the general Markov model. 
Both constructions are capable of capturing standard phylogenetic scenarios where, following speciation, taxa evolve independently (divergence). 
More significantly for the discussion we present here, the results given in \citet{sumner2012algebra} can also be used to model \emph{convergence}, where distinct taxa become more similar over time.
Due to mathematical simplifications present in the restricted case of the K3ST model, this generalization is not available using Bryant's approach. 
We refer to these models as \emph{convergence-divergence models}.

The $n$-taxon process can be thought of as a process on patterns. For the binary symmetric model there are $2^{n}$ possible combinations of states. For example, for $n\!=\!3$ taxa, there are $2^{3}\!=\!8$ possible combinations of states, $000$, $001$, $010$, $011$, $100$, $101$, $110$ and $111$, where the first entry is the state of taxon 1, the second entry is the state of taxon 2 and the third entry is the state of taxon 3. For $4$-state models there are $4^{n}$ possible combinations of states. We call theoretical probabilities of combinations of states occurring \emph{character pattern probabilities} and empirical probabilities \emph{character pattern frequencies}. Character pattern frequencies can, for example, be estimated by frequencies of combinations at sites along nucleotide sequences or from morphological character frequencies.

We wish to determine expressions for each character pattern probability. To do this we start with a probability distribution at the root, for example the stationary distribution. We then have a splitting event that splits the root into two descendant edges (or more if multifurcations are to be considered), followed by Markov processes acting on the edges directly below the root in some epoch of time. A phylogenetic tree or convergence-divergence network can then be constructed from a series of splitting events and Markov processes.

As in the $n$-taxon process, the convergence-divergence model can be thought of as a sequence of Markov processes, each occurring in a separate epoch. 
The epochs are separated by speciation events or the start or end of convergence periods. 
In the $3$-taxon binary character-state case shown in Figure~\ref{fig:fig1} each epoch has an associated $8\times 8$ rate matrix, corresponding to the $2^3=8$ character states possible in this case. 
For instance, in the first epoch taxa $2$ and $3$ are yet to diverge and therefore it is only possible to be in states where $2$ and $3$ are identical. Furthermore, transitions such as from state $011$ to $000$ are possible in a single step. In the third epoch taxa $1$ and $2$ are converging, so states in which $1$ and $2$ are mismatched, e.g. $011$, can transition back to being in a matched state: $001$ or $111$, but the reverse transitions, from $001$ or $111$ to $011$, are not permitted. 
The mathematical details of how the Markov transition matrices are constructed are very interesting from an algebraic standpoint but for the purpose of the paper we do not need these details; the interested reader should consult either \citep{sumner2012algebra} or \citet{mitchell2016distinguishing}.

Convergence-divergence models are not equivalent to any previous phylogenetic network approaches. 
Unlike splits-based methods such as Neighbor-Net \citep{bryant2002neighbornet}, or split decomposition \citep{bandelt1992split}, convergence-divergence models are directed in time and lead to specific predictions regarding character pattern probabilities. 
They are also different from the approaches for implementing maximum likelihood on networks \citep{nakhleh2010evolutionary}. 
As described in Nakhleh's review, a directed network is typically thought of as encoding a set of trees (those displayed by the network). 
The likelihood is then either a mixture model over these trees \citep{jin2006maximum}, or each site is allowed to pick the tree that suits it best. 
Given their stated properties, the convergence-divergence models have different limiting properties to either of these frameworks. 
For instance, if the convergence process is run for a long enough period of time then the taxa that are converging become arbitrarily close.
This is not the case in either the mixture-model setting or for the $n$-taxon process as described by \citet{bryant2009hadamard}.

In his wonderfully entitled paper ``Should phylogenetic models be `trying to fit an elephant' '' \citet{steel2005should} suggested two key points to keep in mind when developing new phylogenetic models. His two points were:
\begin{enumerate}
\item Are they capturing a process that is important biologically?
\item Do they over-fit the data?
\end{enumerate}
With regard to Steel's first point, we believe that convergence-divergence models have potential to be a useful addition to the phylogeneticists' toolkit as there are several biological processes that could be better modeled by considering convergence of taxa rather than regarding networks as a mixture of trees. 
For instance, they might be appropriate for modeling introgression, for example from Neanderthals into humans \citep{green2010draft} or among domesticated and wild plants \citep{ellstrand1999gene}. In the extreme case, despeciation, the loss of unique species, can occur. \citet{rhymer1996extinction} describe how over time introduced species can lead to despeciation of closely related native species through introgression.

In this sense, convergence-divergence models can be thought of as a species-level analogue to the population-level isolation/migration model of \citet{hey2010isolation}. 
\citet{seehausen2008speciation} argued that a loss of diversity can break down ecological boundaries, allowing more opportunities for the exchange of genetic material among previously independent populations, which can in turn lead to convergence. 
\citet{taylor2006speciation} described a case where environmental changes may be resulting in the convergence of three-spined sticklebacks (\textit{G. aculeatus}) in Enos Lake, Vancouver Island. 
\citet{sheppard2008convergence} and \citet{sheppard2011introgression} identified a case in which two species of bacteria, \textit{C. jejuni} and \textit{C. coli}, appear to be in the process of undergoing convergence through horizontal gene transfer. 
A further scenario where we might consider applying convergence-divergence models is for morphological data where selection acts similarly on taxa in different parts of the tree, causing some of the morphological characters to converge \citep{holland2010identifying}. 
One limitation of the convergence-divergence model is that it assumes all characters are independent.

The second point of \citet{steel2005should} is possibly more worrying. 
As noted in \citet{sumner2012algebra} convergence-divergence models have a lot of flexibility: in principle there can be arbitrarily many epochs in which arbitrary groups of taxa can either converge or diverge.
However, under some reasonable restrictions, convergence-divergence models need not necessarily be more parameter rich than trees.
For example, while an $n$-taxon clocklike tree has $n\!-\!1$ height parameters that define the edge lengths and an $n$-taxon non-clocklike tree has $2n-3$ edge parameters, a clocklike convergence-divergence model has at least $n$ epoch length parameters. 
For large $n$, $2n\!-\!3\approx{}2\left(n\!-\!1\right)$ and the number of time parameters will approximately double if we keep the tree assumption but remove the molecular clock assumption. 
Alternatively, we have far more flexibility in the number of parameters if we keep the molecular clock assumption but remove the assumption that once speciation occurs taxa are strictly diverging. We could then increase the number of parameters one at a time until we have optimized the fit. 
Table~\ref{fig:fig4} shows the number of parameters for the three convergence-divergence models on three taxa that we consider in this paper.

Related to the question of whether a model over-fits the data are the fundamental issues of identifiability and distinguishability.  
In this paper we use the term {\em identifiable} in the sense of \citet{allman2008identifying} to mean that there is an (essentially) one-to-one map between the  parameters (e.g. edge lengths or epoch lengths) and the distribution of character pattern probabilities. That is, for every set of parameters there is one possible set of character pattern probabilities and vice versa.
We say that two models with different structural parameters (e.g. a clocklike tree versus a non-clocklike tree) are {\em distinguishable} if there is {\em some} choice of parameters on one of the models that gives character pattern probabilities which cannot arise on the other model.
We say that two models with different structural parameters are {\em distinguishable with respect to a specific set of character pattern probabilities} if it is possible for those character pattern probabilities to arise on one model but not the other.

\citet{sumner2012algebra} raised many unanswered questions regarding both identifiability of model parameters and whether or not the induced character pattern probabilities would be distinguishable from character pattern probabilities arising from tree models. 
These questions were explored in the thesis of \citet{mitchell2016distinguishing}. 
For the $3$-taxon case, \citet{mitchell2016distinguishing} found some scenarios that were neither identifiable nor distinguishable in general from clocklike trees as well as some scenarios that were both identifiable and distinguishable from trees.
 The results presented in \citep{mitchell2016distinguishing} raise some interesting questions about how best to model network evolution and the structure of phylogenetic models in general. 
In this article we explore a simple $3$-taxon scenario and show an example of character pattern probabilities where a non-clocklike tree and convergence-divergence model cannot be distinguished on statistical grounds.

\section{Choosing Between Models}

Many models can be ruled out by an Occam’s Razor argument --- we do not want to consider models that are not identifiable, furthermore if two scenarios produce the same character pattern probabilities then we prefer the scenario with the smaller number of parameters. 
For example, for the binary symmetric model, \citet{mitchell2016distinguishing} showed that the convergence-divergence model for two taxa where the taxa diverge for a period of time and then converge for a period of time is not identifiable and nor is it distinguishable from a model where the taxa simply diverge for a (shorter) period of time. The two-taxon clocklike tree and the two-taxon convergence-divergence model are shown in Figure~\ref{fig:fig3}. 

Furthermore, for the binary symmetric model he argued that if the two-taxon convergence-divergence model is embedded within a larger model then that model will not be identifiable. It will also not be distinguishable from the model created by removing the convergence period.

Restricting to the $3$-taxon case, the following scenarios are distinguishable in general: the clocklike tree (which has two height parameters), the non-clocklike tree (which has three edge parameters), and the convergence-divergence model with convergence between non-sister taxa (which has three epoch length parameters). 
The rightmost subfigure of Figure~\ref{fig:fig1} shows an example of a convergence-divergence model with convergence between non-sister taxa. We do not consider the model in which the two sister taxa converge as it is not identifiable (as mentioned above). We also do not consider convergence-divergence models with more than three epochs, i.e. there is at most one period in which previously diverging taxa experience convergence.
Considering the taxon labeling choices for three taxa, there are three clocklike trees, a single non-clocklike tree and six convergence-divergence models with convergence between non-sister taxa. 

In general we support an approach  where more parameter rich models are only preferred if they offer a substantial improvement to the likelihood \citep{burnham2002model}, so to choose between these ten models given a particular data set (character pattern frequencies)
we fit the parameters (node heights, edge lengths or epoch lengths) using a maximum likelihood approach and then calculate the $AIC = -2 ( \ln{ ( \likelihood{} )}) + 2K$, where $K$ is the number of parameters \citep{burnham2002model}. For small sample sizes the corrected Akaike information criterion can be used. Note that the non-clocklike tree and the non-sister convergence-divergence model have the same number of parameters (three edge or epoch length parameters respectively), whereas the clocklike tree has just two height parameters and so may still be preferred by AIC even if it has a lower likelihood.

\section{Exploring the $3$-Taxon Case}

In one of the earliest papers introducing maximum likelihood to phylogenetics, \citet{felsenstein1981evolutionary} proposed that one test of the molecular clock hypothesis would be to compare the likelihood of models where all edges are free to models with a clock imposed. We extend this idea here by adding an extra possible scenario.
Convergence-divergence models offer another possibility for explaining apparent violations of the molecular clock (i.e. distances that do not obey the three-point condition, where for any three taxa two of the pairwise distances are equal and smaller than the third pairwise distance) --- evolution may not be strictly divergent. 

From the results of \citet{mitchell2016distinguishing}, we know that the clocklike tree, non-clocklike tree and non-sister convergence-divergence model are distinguishable in general. That is, there exist character pattern probabilities that can arise on the non-sister convergence-divergence model that cannot arise on the non-clocklike tree. However, this does not guarantee that the different models will be distinguishable with respect to a particular set of character pattern probabilities.

A question that arises is how to compare the $3$-taxon non-clocklike tree to the $3$-taxon clocklike non-sister convergence-divergence model, as both have three parameters, one more than the $3$-taxon clocklike tree. We wish to know whether there are any circumstances in which we have a choice of a non-clocklike tree or a clocklike convergence-divergence model. To answer this question we explored whether character pattern probabilities that arose on the non-clocklike tree could have also arisen on the clocklike convergence-divergence model and whether character pattern probabilities that arose on the clocklike convergence-divergence model could have also arisen on the non-clocklike tree.
This involved finding algebraic conditions for the probabilities expected under a given model and comparing these conditions for different models.
This approach has been explored by \citet{klaere2012algebraic} and \citet{klare2005stochastic}, who looked at the conditions for the general two-state Markov model on tripod and quartet trees.

Figure~\ref{fig:fig2} shows an example where particular choices of edge lengths for the non-clocklike tree and epoch lengths for the convergence-divergence model give rise to exactly the same set of character pattern probabilities. Additionally, there are two different non-sister convergence-divergence models that both give rise to the same character pattern probabilities.
The same data can be explained either by a violation of the molecular clock, or by supposing convergence between some of the taxa. 

An interesting feature of the three scenarios shown in Figure~\ref{fig:fig2} is that in each scenario taxon $2$ and taxon $3$ are always diverging. Indeed, the path distance from taxon $2$ to taxon $3$ in the non-clocklike tree ($1.43$) is equal to twice the sum of the epoch lengths in both the convergence-divergence models. The distance between taxon $1$ and taxon $2$ is the smallest but this is achieved in different ways in the two models. In the leftmost convergence-divergence model taxon $1$ and taxon $2$ have only been diverging for a height of $0.418+0.013=0.431$ time units, whereas in the rightmost convergence-divergence model these taxa have been diverging for a height of $0.033+0.528=0.561$ time units, but then have subsequently converged for $0.154$ time units.

Suppose we fix a $3$-taxon non-clocklike tree.
We wish to determine whether an equivalent set of epoch lengths on a $3$-taxon clocklike non-sister convergence-divergence model can be found.
That is, we want to find whether there is a set of epoch lengths on the convergence-divergence model, such that all character pattern probabilities for the convergence-divergence model are equal to those for the non-clocklike tree.

We start by finding expressions for the character pattern probabilities in terms of the edge lengths for the non-clocklike tree and in terms of the epoch lengths for the convergence-divergence model. By equating the character pattern probabilities for the non-clocklike tree and the convergence-divergence model, we can find expressions for the epoch lengths of the convergence-divergence model in terms of the edge lengths of the non-clocklike tree. Placing no restrictions on the set of edge lengths for the non-clocklike tree, there will always be one edge length that is less than or equal to the other two edge lengths. If we choose the taxon on this edge to be the taxon on the convergence-divergence model that is both in the cherry and involved in convergence then a set of epoch lengths on the convergence-divergence model can always be found that preserves the character pattern probabilities.

We are free to choose between the two choices of where we will place the remaining two taxa on the convergence-divergence model. The convergence-divergence models will always come in pairs, as seen in the example in Figure~\ref{fig:fig2}. Note however, that the epoch lengths on the two convergence-divergence models are not generally equal, but the character pattern probabilities are.

As a consequence, we are always free to choose between two taxon labellings on the convergence-divergence model. We can choose a convergence-divergence model where two taxa split well after the root, with only a short period of convergence between two non-sister taxa at the leaves, as shown in the diagram at the bottom left of Figure~\ref{fig:fig2}. Alternatively, we can choose a convergence-divergence model where the three taxa split from each other soon after the root, with a long period of convergence between the two non-sister taxa at the leaves, as shown in the diagram at the bottom right of Figure~\ref{fig:fig2}. These two convergence-divergence models will have identical log likelihoods. We may instead wish to choose the model that seems more biologically realistic.

We conducted a simulation study to demonstrate that situations such as those shown in Figure~\ref{fig:fig2} always arise on the $3$-taxon non-clocklike tree. We began with a non-clocklike tree and randomly selected three edge lengths. The character pattern frequencies on this tree were determined and we then checked to see if we could find a non-sister convergence-divergence model that achieved the same character pattern frequencies.

Of $1000$ random non-clocklike trees, all $1000$ gave character pattern frequencies that could be matched on a convergence-divergence model. 
Although it is not provided here, it can be proven that a matched convergence-divergence model exists for all choices of edges on the non-clocklike tree.

Note that in \citet{mitchell2016distinguishing} the potential for character pattern probabilities that could have arisen on the non-clocklike tree, but not on the non-sister convergence-divergence network, was left open. It is now known that these character pattern probabilities do not occur and that character pattern probabilities that arose on a non-clocklike tree can always be matched to character pattern probabilities that arose on a non-sister convergence-divergence network.

We then did a similar simulation, but beginning instead with random choices of epoch lengths on the non-sister convergence-divergence model. The character pattern frequencies on the model were determined and we then checked to see if we could find a non-clocklike tree that achieved the same character pattern frequencies. Of $1000$ random models $873$ gave character pattern frequencies that could be fit equally well on a non-clocklike tree.

Suppose character pattern probabilities arose on the clocklike tree. There will always be character pattern probabilities that could have arisen on either the non-clocklike tree or the non-sister convergence-divergence network. Now suppose that character pattern probabilities arose on the non-clocklike tree. There will always be character pattern probabilities that could have arisen on the convergence-divergence network. Finally, suppose that character pattern probabilities arose on the convergence-divergence network. There will \emph{sometimes} be character pattern probabilities that could have arisen on the non-clocklike tree. Given that character pattern probabilities arose on either the non-clocklike tree or the convergence-divergence network, they could have only arisen on the clocklike tree if some of the epoch lengths on the corresponding convergence-divergence model were 0 or if two of the edges on the non-clocklike tree were equal to each other. 
This is summarized in Table~\ref{fig:fig4}.

\section{Discussion}


Under the assumption of treelike evolution, the number of structural parameters needed is determined by the number of species under consideration (e.g. in a non-clocklike binary tree with $n$ taxa, we have $2n-3$ edges).
Thus phylogeneticists usually need not think about the number of structural parameters required.
However, when we remove the tree assumption we have to think more carefully about variable selection \citep{holland2013rise}.

It is increasingly apparent that there are many biologically reasonable causes of non-treelike evolution, so it seems clear that we should consider a broader range of models.
In many cases it will be possible to choose between competing scenarios on the basis of AIC, however, the example we give here shows that this will not always be the case.
In the scenario presented here we show two competing biologically reasonable scenarios that give identical likelihoods and have the same number of parameters. Which should we prefer? In the end it depends if we believe more in a molecular clock or if we believe more strongly in a divergence-only model.
Sometimes we may be able to bring extra information to the problem. For example, to choose between two pairs of models we might know that two species are sympatric and that there is some opportunity for gene-flow whereas two other species are not. Note, however, that we don't have a \textit{statistical} way to choose between models and cannot escape the need for knowledge of the underlying biology/geography.

As discussed earlier, \citet{steel2005should} raised two important issues regarding model fit. We want a model that has both enough parameters to be biologically realistic and few enough parameters to not over-fit the data. In the past, phylogeneticists have often rejected models with the molecular clock hypothesis due to their tendency to not have enough parameters to be biologically realistic. In order to introduce additional parameters to the model phylogeneticists commonly drop the molecular clock assumption for non-clocklike trees. Removing the molecular clock assumption approximately doubles the number of edge parameters for large numbers of taxa, which can result in over-fitting the data. If both a clocklike tree under-fits the data and a non-clocklike tree over-fits the data then a model with an intermediate number of edge or epoch parameters may be more appropriate. The relaxed clock models of \citet{drummond2006relaxed} have been popular and effective in terms of giving flexibility to fit data without requiring so many extra parameters. Our clocklike convergence-divergence models provide an alternative approach to give more freedom in the number of parameters. Any number of extra parameters beyond those on clocklike trees can be introduced. Extra parameters can be added one at a time until there is no statistically significant improvement to the fit of the model to the data, as judged by the AIC. One must take care in the order of introduction of convergence epochs as there are many possible scenarios that could be modeled with convergence-divergence models.

Future work may involve exploring more than three taxa, in particular four taxa, as well as examining more biologically realistic Markov models than the binary symmetric model. The binary symmetric model was chosen to clearly illustrate our salient points, however in principle the same mathematical techniques can be applied to the general Markov model on any number of states.




\section{Funding}

This work was supported by the Australian Postgraduate Award, Australian Research Council awards FT100100031 and DP150100088, and the US National Institutes of Health grant R01 GM117590, awarded under the Joint DMS/NIGMS Initiative to Support Research at the Interface of the Biological and Mathematical Sciences.



\bibliographystyle{sysbio}
\bibliography{tripod}

\newpage

\begin{table}
\caption {The number of parameters and overlap with other models. 
The third column indicates how the models overlap. 
} \label{fig:fig4} 
\begin{center}
\begin{tabular}{ >{\centering\arraybackslash}m{3.5in}  >{\centering\arraybackslash}m{.75in} >{\centering\arraybackslash}m{.75in}}
\toprule[1.5pt]
{Model} & {Parameters} & {Overlap}\\ 
\midrule
Clocklike Tree (a) & 2 & -\\
\midrule
Non-Clocklike Tree (b) & 3 & a\\
\midrule
Non-Sister Convergence-Divergence Model (c) & 3 & a,b\\
\bottomrule[1.25pt]
\end {tabular}
\end{center}
\end{table}

\begin{figure}
	\centering
				\begin{tikzpicture}
               	\centering
				\draw[-] (1.71,0) -- (1,-2) node [below] {1};
				\draw[-] (1.71,0) -- (2.42,-2) node [below] {2};
				\draw[-] (5.84,0) -- (4.42,-4);
				\draw[-] (5.84,0) -- (7.26,-4);
                \draw[bend left] (4.42,-4) to (5.17,-4.25) node [below] {1};
				\draw[bend right] (7.26,-4) to (6.51,-4.25) node [below] {2};
				\begin{pgfonlayer}{background}
				\draw[-,dashed] (0,0) -- (8.26,0);
				\draw[-,dashed] (0,-2) -- (8.26,-2);
				\draw[-,dashed] (0,-4) -- (8.26,-4);
				\draw[-,dashed] (0,-4.25) -- (8.26,-4.25);
				\end{pgfonlayer}
			\end{tikzpicture}
                    \caption{A two-taxon clocklike tree on the left and a two-taxon convergence-divergence model on the right, to scale in the vertical direction. The length of the convergence epoch is chosen so that the character pattern probabilities of the tree and convergence-divergence model are identical. The convergence-divergence model is not identifiable, nor is it distinguishable from the tree.}
		\label{fig:fig3}
\end{figure}
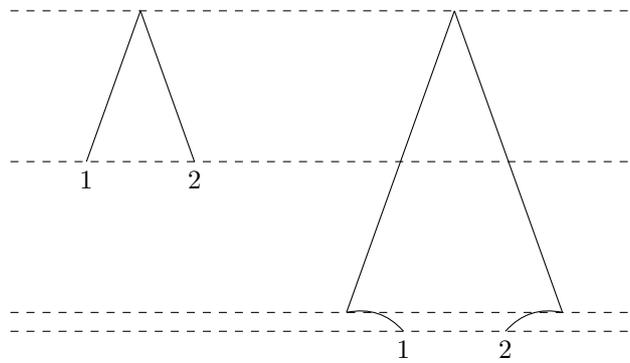

\begin{figure}
	\centering
			\begin{subfigure}{\textwidth}
            	\centering
				\begin{tikzpicture}
               	\centering
				\draw[-] (1.71,0) -- (1,-2) node [below] {1};
				\draw[-] (1.71,0) -- (2.42,-2) node [below] {2 3};
				\draw[-,transform canvas={xshift=3pt}] (1.71,0) -- (2.42,-2);
				\draw[-] (5.84,0) -- (4.42,-4) node [below] {1};
				\draw[-] (5.84,0) -- (6.55,-2);
				\draw[-] (6.55,-2) -- (5.84,-4) node [below] {2};
				\draw[-,transform canvas={xshift=3pt}] (5.84,0) -- (7.26,-4) node [below] {3};
				\draw[-] (10.68,0) -- (9.26,-4);
				\draw[-] (10.68,0) -- (11.39,-2);
				\draw[-] (11.39,-2) -- (10.68,-4);
				\draw[bend left=15] (9.26,-4) to (9.74,-6) node [below] {1};
				\draw[bend right=15] (10.68,-4) to (10.20,-6) node [below] {2};
				\draw[-,transform canvas={xshift=3pt}] (10.68,0) -- (12.81,-6)  node [below] {3};
				\begin{pgfonlayer}{background}
				\draw[-,dashed] (0,0) -- (13.81,0);
				\draw[-,dashed] (0,-2) -- (13.81,-2);
				\draw[-,dashed] (0,-4) -- (13.81,-4);
				\draw[-,dashed] (0,-6) -- (13.81,-6);
				\end{pgfonlayer}
			\end{tikzpicture}
			\end{subfigure}
\par\bigskip
    	\begin{subfigure}{0.32\textwidth}
        	\centering
			\begin{tikzpicture}[->,node distance=2.5cm,thick,main node/.style={circle,draw}]
			\node[main node,fill=black!10!] (1) {110};
			\node[main node,fill=black!40!] (2) [below left of=1] {100};
			\node[main node,fill=black!40!] (3) [below right of=1] {111};
			\node[main node,fill=black!10!] (4) [below of=1] {010};
			\node[main node,fill=black!40!] (5) [below of=2] {000};
			\node[main node,fill=black!40!] (6) [below of=3] {011};
			\node[main node,fill=black!10!] (7) [below right of=2] {101};
			\node[main node,fill=black!10!] (8) [below right of=5] {001};
			\draw [->] (1) to (4);
			\draw [->] (4) to (1);
			\draw [->] (2) to (5);
			\draw [->] (5) to (2);
			\draw [->] (3) to (6);
			\draw [->] (6) to (3);
			\draw [->] (7) to (8);
			\draw [->] (8) to (7);
			\draw [->] (2) to (3);
			\draw [->] (3) to (2);
			\draw [->] (5) to (6);
			\draw [->] (6) to (5);
			\draw[dotted] [->] (1) to (2);
			\draw[dotted] [->] (1) to (3);
			\draw[dotted] [->] (4) to (5);
			\draw[dotted] [->] (4) to (6);
			\draw[dotted] [->] (7) to (2);
			\draw[dotted] [->] (7) to (3);
			\draw[dotted] [->] (8) to (5);
			\draw[dotted] [->] (8) to (6);		
		\end{tikzpicture}
		\end{subfigure}
    	\begin{subfigure}{0.32\textwidth}
        	\centering
			\begin{tikzpicture}[->,node distance=2.5cm,thick,main node/.style={circle,draw}]
			\node[main node,fill=black!40!] (1) {110};
			\node[main node,fill=black!40!] (2) [below left of=1] {100};
			\node[main node,fill=black!40!] (3) [below right of=1] {111};
			\node[main node,fill=black!40!] (4) [below of=1] {010};
			\node[main node,fill=black!40!] (5) [below of=2] {000};
			\node[main node,fill=black!40!] (6) [below of=3] {011};
			\node[main node,fill=black!40!] (7) [below right of=2] {101};
			\node[main node,fill=black!40!] (8) [below right of=5] {001};
			\draw [->] (1) to (4);
			\draw [->] (4) to (1);
			\draw [->] (2) to (5);
			\draw [->] (5) to (2);
			\draw [->] (3) to (6);
			\draw [->] (6) to (3);
			\draw [->] (7) to (8);
			\draw [->] (8) to (7);
			\draw [->] (2) to (3);
			\draw [->] (3) to (2);
			\draw [->] (5) to (6);
			\draw [->] (6) to (5);
			\draw [->] (1) to (2);
			\draw [->] (1) to (3);
			\draw [->] (4) to (5);
			\draw [->] (4) to (6);
			\draw [->] (7) to (2);
			\draw [->] (7) to (3);
			\draw [->] (8) to (5);
			\draw [->] (8) to (6);		
		\end{tikzpicture}
		\end{subfigure}
    	\begin{subfigure}{0.32\textwidth}
        	\centering
			\begin{tikzpicture}[->,node distance=2.5cm,thick,main node/.style={circle,draw}]
			\node[main node,fill=black!40!] (1) {110};
			\node[main node,fill=black!40!] (2) [below left of=1] {100};
			\node[main node,fill=black!40!] (3) [below right of=1] {111};
			\node[main node,fill=black!40!] (4) [below of=1] {010};
			\node[main node,fill=black!40!] (5) [below of=2] {000};
			\node[main node,fill=black!40!] (6) [below of=3] {011};
			\node[main node,fill=black!40!] (7) [below right of=2] {101};
			\node[main node,fill=black!40!] (8) [below right of=5] {001};
			\draw [->] (1) to (3);
			\draw [->] (1) to (5);
			\draw [->] (2) to (7);
			\draw [->] (3) to (1);
			\draw [->] (3) to (8);
			\draw [->] (4) to (6);
			\draw [->] (5) to (1);
			\draw [->] (5) to (8);
			\draw [->] (6) to (4);
			\draw [->] (7) to (2);
			\draw [->] (8) to (3);
			\draw [->] (8) to (5);
			\draw[dotted] [->] (2) to (1);
			\draw[dotted] [->] (2) to (5);
			\draw[dotted] [->] (4) to (1);
			\draw[dotted] [->] (4) to (5);
			\draw[dotted] [->] (6) to (3);
			\draw[dotted] [->] (6) to (8);
			\draw[dotted] [->] (7) to (3);
			\draw[dotted] [->] (7) to (8);		
		\end{tikzpicture}
        \end{subfigure}
        \caption{Three epochs of a $3$-taxon process. (Top) Divergence of taxa is represented by straight lines emanating from a node, while convergence in the third epoch is represented by curved lines. (Bottom) State transition diagrams for each epoch. Light gray nodes and arrows in the left-most diagram indicate states which it is not possible to be in as taxa $j$ and $k$ are forced to be identical. In the middle diagram all three taxa are diverging from each other and any single step transition is possible. In the right-most diagram solid lines indicate ``regular'' mutations, while dotted lines indicate ``correction'' transitions responsible for convergence.}
		\label{fig:fig1}
\end{figure}
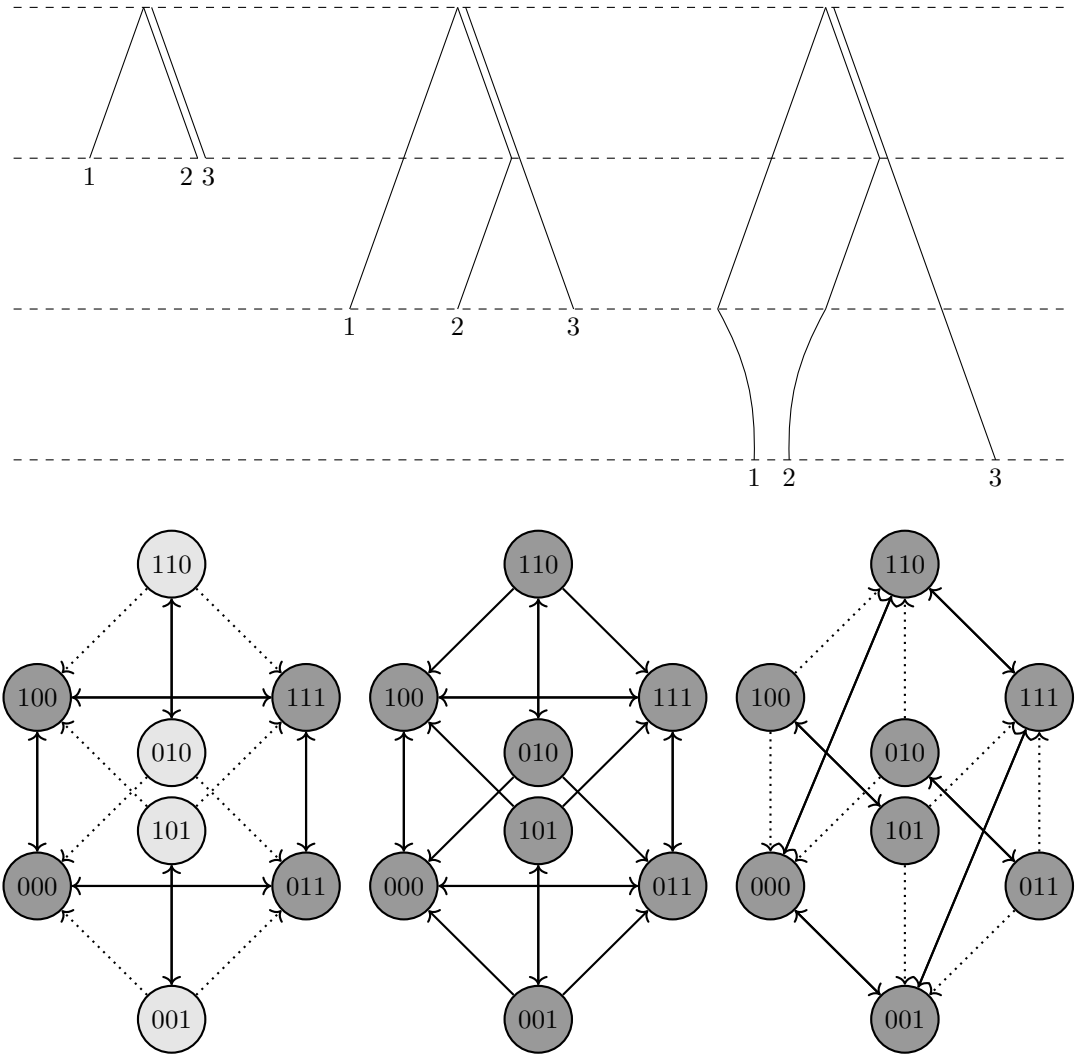

\begin{figure}
\centering
\begin{subfigure}{\textwidth}
\centering
\begin{tikzpicture}
\draw[-] node [left] {3} (0,0) -- (9.66,0) node [midway, above=10pt] {0.966};
\draw[-] (9.66,0) -- (11.65,3.45) node [above right] {1} node [midway, right=10pt] {0.398};
\draw[-] (9.66,0) -- (11.98,-4.02) node [below right] {2} node [midway, right=10pt] {0.464};
\end{tikzpicture}
\end{subfigure}
\par\bigskip
\begin{subfigure}{0.35\textwidth}
\centering
\begin{tikzpicture}[xscale=0.35]
\draw[{Latex[scale=1.5]}-{Latex[scale=1.5]}] (0,0) -- (0,-2.84) node [midway, left=10pt] {0.284};
\draw[{Latex[scale=1.5]}-{Latex[scale=1.5]}] (0,-2.84) -- (0,-7.02) node [midway, left=10pt] {0.418};
\draw[{Latex[scale=1.0]}-{Latex[scale=1.0]}] (0,-7.02) -- (0,-7.15) node [midway, left=10pt] {0.013};
\draw[-] (8.02,0) -- (1,-7.02);
\draw[-] (8.02,0) -- (10.86,-2.84);
\draw[-] (10.86,-2.84) -- (6.68,-7.02);
\draw[bend left] (1,-7.02) to (1.39,-7.15) node [below] {3};
\draw[bend right] (6.68,-7.02) to (6.29,-7.15) node [below] {1};
\draw[-,transform canvas={xshift=3pt}] (8.02,0) -- (15.17,-7.15) node [below] {2};
\begin{pgfonlayer}{background}
\draw[-,dashed] (0,0) -- (16.17,0);
\draw[-,dashed] (0,-2.84) -- (16.17,-2.84);
\draw[-,dashed] (0,-7.02) -- (16.17,-7.02);
\draw[-,dashed] (0,-7.15) -- (16.17,-7.15);
\end{pgfonlayer}
\end{tikzpicture}
\end{subfigure}
\hspace{6em}
\begin{subfigure}{0.4\textwidth}
\centering
\begin{tikzpicture}[xscale=0.35]
\draw[{Latex[scale=1.0]}-{Latex[scale=1.0]}] (0,0) -- (0,-0.33) node [midway, left=10pt] {0.033};
\draw[{Latex[scale=1.5]}-{Latex[scale=1.5]}] (0,-0.33) -- (0,-5.61) node [midway, left=10pt] {0.528};
\draw[{Latex[scale=1.5]}-{Latex[scale=1.5]}] (0,-5.61) -- (0,-7.15) node [midway, left=10pt] {0.154};
\draw[-] (6.61,0) -- (1,-5.61);
\draw[-] (6.61,0) -- (6.94,-0.33);
\draw[-] (6.94,-0.33) -- (5.56,-5.61);
\draw[bend left=25] (1,-5.61) to (1.62,-7.15) node [below] {2};
\draw[bend right=25] (5.56,-5.61) to (4.94,-7.15) node [below] {1};
\draw[-,transform canvas={xshift=3pt}] (6.61,0) -- (13.76,-7.15) node [below] {3};
\begin{pgfonlayer}{background}
\draw[-,dashed] (0,0) -- (14.76,0);
\draw[-,dashed] (0,-0.33) -- (14.76,-0.33);
\draw[-,dashed] (0,-5.61) -- (14.76,-5.61);
\draw[-,dashed] (0,-7.15) -- (14.76,-7.15);
\end{pgfonlayer}
\end{tikzpicture}
\end{subfigure}
\caption{Three biologically different scenarios that are indistinguishable based on the  character pattern probabilities that they induce. All edge and epoch lengths are drawn to scale.}
\label{fig:fig2}
\end{figure}
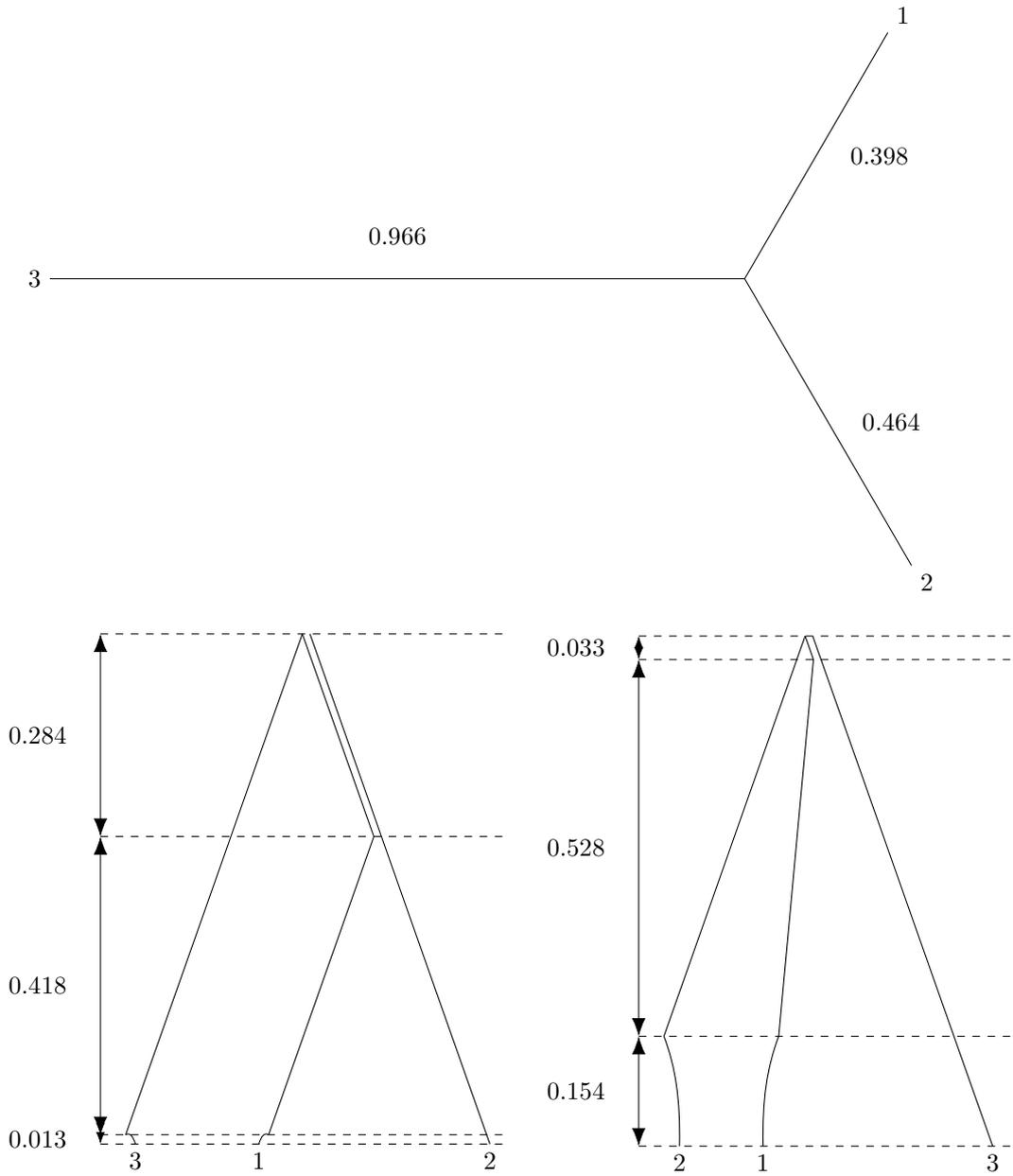

\end{document}